\documentclass[12pt,preprint]{aastex}
\usepackage[dvips]{color}
\newcommand\nion[2]{#1\,\lowercase{{\sc #2}}}

\def\teff{\mbox{T$_{\rm eff}$}}
\def\logg{\rm log~$g$}
\def\tphot{\mbox{T$_{\rm phot}$}}
\def\BmV0{\mbox{$(B-V)^{\rm 0}$}}
\def\VmK0{\mbox{$(V-K)^{\rm 0}$}}
\def\MV0{\mbox{$M_{\rm V}^{\rm 0}$}}

\def\mhr{[m/H]$_{\rm RAVE}$}

\shorttitle{}
\begin{document}

\title{The RAVE Survey:  Rich in Very Metal-Poor Stars
\footnote{Based in part on observations collected at the
European Organization for Astronomical Research in the Southern
Hmeisphere, Chile, in the framework of proposals 081.B-0900
and 080.B-0927}}
\author{
Jon~P.~Fulbright\altaffilmark{1},
Rosemary~F.~G.~Wyse\altaffilmark{1},
Gregory~R.~Ruchti\altaffilmark{1},
G.~F.~Gilmore\altaffilmark{2},
Eva Grebel\altaffilmark{3},
O.~Bienaym\'{e}\altaffilmark{4}, J.~Binney\altaffilmark{5},
J.~Bland-Hawthorn\altaffilmark{6}, 
R.~Campbell\altaffilmark{7}, K.~C.~Freeman\altaffilmark{8},
B.~K.~Gibson\altaffilmark{9}, 
A.~Helmi\altaffilmark{10}, U.~Munari\altaffilmark{11},
J.~F.~Navarro\altaffilmark{12}, Q.~A.~Parker\altaffilmark{13},
W.~Reid\altaffilmark{13},
G.~M.~Seabroke\altaffilmark{14}, A.~Siebert\altaffilmark{4},
A.~Siviero\altaffilmark{12,15}, M.~Steinmetz\altaffilmark{15},
F.~G.~Watson\altaffilmark{6}, M.~Williams\altaffilmark{15}, T.~Zwitter\altaffilmark{16}
}

\altaffiltext{1}{Johns Hopkins University, Department of Physics \&
  Astronomy, 366 Bloomberg Center, 3400 North Charles Street,
  Baltimore, MD 21218, USA} 
\altaffiltext{2}{Institute of Astronomy, University of
  Cambridge, Madingley Road, Cambridge CB3 0HA, UK}
\altaffiltext{3}{Astronomisches Rechen-Institut, Zentrum f\"ur
  Astronomie der Universit\"at Heidelberg, M\"onchhofstr.\ 12--14,
  D-69120 Heidelberg, Germany}
\altaffiltext{4}{Observatoire de Strasbourg, 11 Rue de
  L'Universit\'{e}, 67000 Strasbourg, France} 
\altaffiltext{5}{Rudolf Pierls Center for Theoretical Physics,
  University of Oxford, 1 Keble Road, Oxford OX1 3NP, UK}
\altaffiltext{6}{Anglo-Australian Observatory, P.O. Box 296, Epping,
  NSW 1710, Australia}
\altaffiltext{7}{Western Kentucky University, Bowling Green, Kentucky,
  USA}
\altaffiltext{8}{RSAA Australian National University, Mount Stromlo
  Observatory, Cotter Road, Weston Creek, Canberra, ACT 72611,
  Australia}
\altaffiltext{9}{Jeremiah Horrocks Institute for Astrophysics \&
  Super-computing, University of Central Lancashire, Preston, UK}
\altaffiltext{10}{Kapteyn Astronomical Institute, University of
  Groningen, Postbus 800, 9700 AV Groningen, Netherlands}
\altaffiltext{11}{INAF Osservatorio Astronomico di
  Padova, Via dell'Osservatorio 8, Asiago I-36012, Italy}
\altaffiltext{12}{University of Victoria, P.O. Box 3055, Station CSC,
  Victoria, BC V8W 3P6, Canada} 
\altaffiltext{13}{Macquarie University, Sydney, NSW 2109, Australia}
\altaffiltext{14}{Mullard Space Science Laboratory, University College
  London, Holmbury St Mary, Dorking, RH5 6NT, UK}
\altaffiltext{15}{Astrophysikalisches Institut Potsdam,
  An der Sternwarte 16, D-14482 Potsdam, Germany}
\altaffiltext{16}{Faculty of Mathematics and Physics, University of
  Ljubljana, Jadranska 19, Ljubljana, Slovenia}

\begin{abstract}
  Very metal-poor stars are of obvious importance for many problems in
  chemical evolution, star formation, and galaxy evolution. Finding
  complete samples of such stars which are also bright enough to allow
  high-precision individual analyses is of considerable interest. We
  demonstrate here that stars with iron abundances [Fe/H] $ < -2$~dex,
  and down to below $-4$~dex, can be efficiently identified within the
  Radial Velocity Experiment (RAVE) survey of bright stars, without
  requiring additional confirmatory observations.  We determine a
  calibration of the equivalent width of the Calcium triplet lines
  measured from the RAVE spectra onto true [Fe/H], using high spectral
  resolution data for a subset of the stars.  These RAVE iron
  abundances are accurate enough to obviate the need for confirmatory
  higher-resolution spectroscopy. Our initial study has identified 631
  stars with [Fe/H] $ \leq -2$, from a RAVE database containing
  approximately 200,000 stars.  This RAVE-based sample is complete for stars
  with [Fe/H] $\la -2.5$, allowing statistical sample analysis.  We 
identify three stars with [Fe/H] $ \la -4$.  Of these, one was already known to be `ultra metal-poor', one is a known carbon-enhanced metal-poor star, but we obtain 
[Fe/H]=$-4.0$, rather than the published [Fe/H]=$-3.3$, and derive
[C/Fe] = $+0.9$, and [N/Fe] = $+3.2$, and the third is at the limit of our S/N.  RAVE observations are on-going
and should prove to be a rich source of bright, easily studied, very
metal-poor stars.

\end{abstract}


\section{Introduction}

Very metal-poor (VMP) stars, defined conventionally as having [Fe/H]
$< -2$ \citep{bee05}, are of interest in many outstanding problems in
star formation and galaxy evolution.  For example, the detailed shape
of the metal-poor tail of the metallicity distribution of field stars
of the Milky Way can distinguish different models of chemical
evolution, constraining gas flows, the nature of the progenitor
systems in which stars form, and pregalactic enrichment \citep[e.g.][]{pra09}. Elemental abundance patterns of VMP stars are
sensitive to the high-redshift stellar Initial Mass Function,
star-formation process, and possibly identify the sources responsible
for reionization of the Universe \citep{bro04}.

Detailed elemental abundances of metal-poor stars require
high-resolution high-quality spectroscopy.  Thus apparently bright
targets with high probability of being truly very metal-poor are
desirable.  We demonstrate in this paper that the RAVE survey is an
ideal source of bright metal-poor stars; these
can be identified directly from the RAVE data, with no need for
confirmatory higher-resolution spectroscopy. We derive the calibration
by which the RAVE parameters may be placed on a true iron abundance
scale, valid for [Fe/H] values as metal-poor as $-4$~dex, into the
regime of `ultra metal-poor stars' in the nomenclature of Beers \&
Christlieb (2005).

\section{Identifying Very Metal-Poor Stars in the RAVE Survey}

\subsection{The RAVE Survey}

The Radial Velocity Experiment (RAVE) \citep{ste06, zwi08} is a
spectroscopic survey of apparently bright stars, $I \lesssim
13$, with some 350,000 stars observed to date.  RAVE uses the
six-degree-field (6dF) multi-object, fibre-fed spectrograph on the
UK~Schmidt telescope.  Spectra of $\sim
100$ targets are obtained simultaneously. The dominant target
population belongs to the thin disk, but includes many thick disk
and halo giant stars. The spectra are centered on the infrared Calcium
triplet, cover the wavelength range region $\lambda\lambda 8410 -
8795$\AA\, and have resolving power $R \sim 7500$. This is
significantly higher than that which is typically employed in surveys
to find metal-poor candidates for later confirmation -- for example,
the Hamburg/ESO survey \citep{chr04} is based on objective-prism
spectra with $R \sim 400$ at the Ca II K line at 3934\AA. We show
here that RAVE has the resolution to determine the metal-poor nature of
a star directly from the survey spectra, without the need for
follow-up confirmation, providing ideal targets for later detailed
analysis.

The RAVE spectral analysis pipeline provides estimates for each star
of the line-of-sight velocity and the stellar parameters (T$_{\rm
  eff}$, $\log{ g}$, [m/H], and $v\sin{i}$) through $\chi^2$ fits to a
grid of synthetic spectra\footnote{The quantity [m/H] here designates
  the total amount of metals with respect to the solar value. We will
  use [Fe/H] to refer to only the iron content.}. RAVE pipeline
metallicities ([m/H]$_{{\rm RAVE}}$) have uncertainty of $\sim
0.2$~dex, derived from an external calibration of high-quality spectra
of predominantly metal-rich stars, which dominate the sample
(Zwitter et al.~2008).  The grid of synthetic spectra utilized extends
to [m/H] $= -2.5$, requiring additional analysis below this limit: we 
provide that here.

\subsection{Isolating candidate Very Metal Poor Stars in RAVE}

The RAVE survey is designed
primarily to obtain excellent line-of-sight velocities.  This has been
achieved, with accuracy and precision of a few km/s for the vast majority
of the stars (Steinmetz et al.~2006).  A large subset of the spectra
has sufficiently high signal-to-noise ratio for the RAVE pipeline also to
provide good quality estimates of the values of stellar parameters
(Zwitter et al.~2008).  At the start of the present investigation, the
RAVE (internal) database contained over 147,000 spectra from which
stellar parameters were derived. How do we identify VMP stars among
these?  Stars with true metallicity near or below the lower limit of
the synthetic spectra in the RAVE pipeline
($-2.5$~dex) will be assigned metallicities close to that lower
limit. A metallicity cut in the RAVE database of, say, 3$\sigma$  above
the lower bound, \mhr $ \leq -1.8$, should then isolate probable VMP
stars.  Adopting this cut  and isolating  only normal stars  gives a sample of 622 potential metal-poor stars.

\begin{figure}
\includegraphics[angle=270,width=6truein]{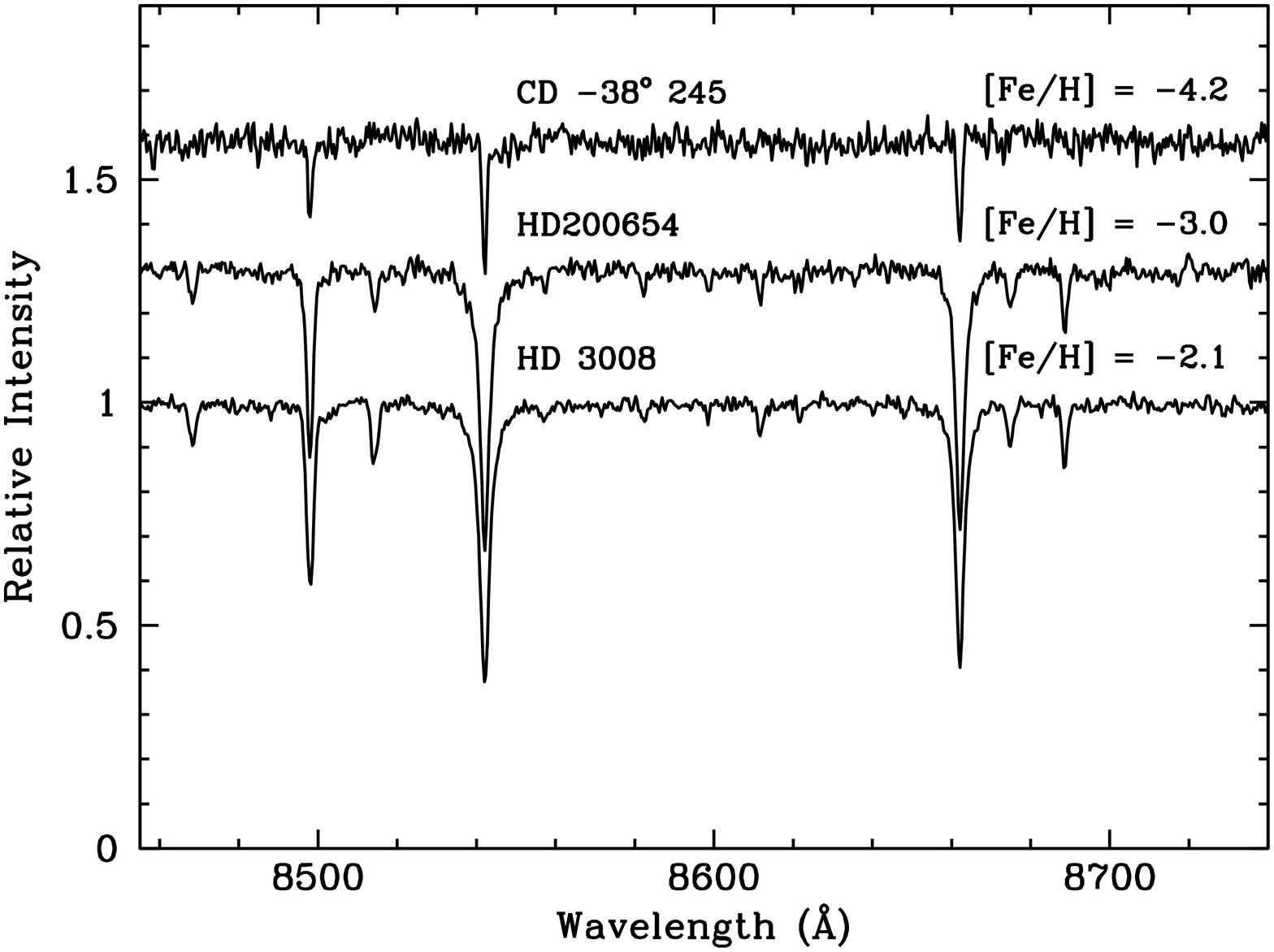}
\caption{RAVE spectra of three metal-poor stars, with our derived iron abundance estimates (consistent with previous estimates in the literature). Note the dominance of the Ca II Triplet lines, and the growing strength of Fe~I and other species as [Fe/H] increases.}
\label{3stars}
\end{figure}

Furthermore, extremely metal-poor stars will be erroneously assigned
to significantly higher metallicities by the RAVE pipeline.  The RAVE spectrum
of a typical VMP star is nearly featureless, with only the Ca Triplet
lines being measurable, as shown in Figure~\ref{3stars}.  There exists a degeneracy between the stellar
parameters and metallicity, in that the strengths of the Ca Triplet
lines can be equal for both a lower-metallicity cooler giant and a
higher-metallicity hotter turn-off star.  A temperature difference of
100~K corresponds to a \mhr{} difference of about $0.1$ dex,
assuming the parameters of the star still lie along an appropriately
metal-poor old evolutionary track.  Hence stars for which the RAVE
\teff{} value is significantly hotter than the value that is derived
by other indicators are likely to be stars whose real [m/H] values are
significantly lower than \mhr{}. This expectation is confirmed by
identification in the RAVE database of several known VMP stars,
incorrectly allocated \mhr $> -1.8$, including CD~$-38^\circ245$, with 
[Fe/H]~$= -4.2$ \citep{cay04} and CS~29502-092 with [Fe/H]~$=~-2.76$
\citep{aok07}.  As anticipated, known VMP stars with anomalously high
\mhr{} values also have anomalously high estimated stellar effective
temperatures from the RAVE pipeline.

Temperature anomalies may be identified using photometric
temperatures.  Photometry from the 2MASS survey \citep{cut03} is
available for the RAVE  sample, so that an effective
temperature, \tphot, may be calculated using the (de-reddened)
color-\tphot{} relationship of \citet{alo99}, with due care to use the 2-color relations to check for 
cool binary companions. The mean value of E(B-V)
for our candidate metal-poor stars is 0.11~mag (median 0.06~mag).
We assume conservatively that the stars of interest are beyond the
reddening layer, since their estimated distances based on the stellar
parameters are typically several hundred parsecs to a
few kiloparsecs from the Sun.

These \tphot\ values allow us to adopt the following  
modified criterion to define a candidate VMP star:

$$ {\rm [m/H]_{RAVE}} - 0.1(({\rm T_{\rm RAVE}} - {\rm T_{\rm phot}})/(100 {\rm K})) < -1.8. $$

Application of this selection criterion to the RAVE database of non-variable stars within the appropriate temperature range to be halo turnoff/red giant stars (7000 $ < T_{{\rm RAVE}} <$  3800~K) 
yields 320 stars in addition
to those selected by the straight cut \mhr $ \leq -1.8$.  The total
number of stars we selected for further analysis of their RAVE
spectrum, based on the above criteria, is then 942.

\section{Derivation of [Fe/H] Abundances}

\subsection{Equivalent-Width Analysis of RAVE Spectra}

For each candidate VMP star we measure the equivalent width (EW) of
the Ca Triplet and Fe~I lines from the RAVE spectrum.\footnote{At
least one of the Ca II lines is too strong in the majority of the
candidates. All three Ca II lines could be used in 132 stars; for
these the mean dispersion is 0.26~dex, consistent with our
final estimated error.} These EW values are then the input for an
abundance analysis using the MOOG program \citep{sne73}, assuming LTE
and 1-D, plane-parallel Kurucz model atmospheres.\footnote{The most
recent versions of the Kurucz atmospheres can be found at
http://kurucz.harvard.edu/.}  We adopted the NLTE corrections of
\citet{sta10} for all our stars (in the end significant only for stars with [Fe/H] $<
-2.5$).

The estimates of stellar gravity from the RAVE pipeline analysis
identify most VMP candidates as giants: 86\% of these stars have RAVE
\logg~$<~3$, with the mean RAVE \logg{} value being $1.0 \pm 0.9$.
Stellar gravity values from the RAVE pipeline are accurate to better
than 0.5~dex for the typical metal-rich stars in the RAVE database
(Zwitter et al.~2008). For metal-poor stars the lower metallicity
cut-off of the synthetic spectrum grid biases the pipeline analysis to
overestimate, systematically, both \teff{} and \logg{} values, so the
actual fraction of true giants in the VMP sample should be even
higher. Indeed, there were only four stars whose location on a reduced
proper-motion diagram indicated they were dwarfs.  For these we adopt
\logg~$=4.5$.  The stellar surface gravity for the remaining stars was
obtained by fitting to an old (12~Gyr) Yonsei-Yale isochrone
\citep{yy} with a metallicity matching the current metallicity
estimate of the star (the procedure is iterated).  This gravity value
was adopted in the calculation of the model atmosphere (while the Yonsi-Yale 
isochrones do not extend beyond the RGB, $< 0.1$~dex error in final
abundance results should a star be actually on the Asymptotic Giant
Branch). Our analysis was iterated until the derived (from CaT)
metallicity did not differ by more than 0.1~dex from the value used in
calculating the model atmosphere and isochrone fit.  We were able to
derive Ca abundances for 771 stars out of our 942 star initial
selection from RAVE: these 771 stars form our VMP candidate sample.

The next step is to provide an external calibration of our abundance
scale, derived from high signal-to-noise echelle spectra.

\subsection {Metallicity Calibration {\em via}   Echelle Observations}

We obtained echelle spectroscopic data for 112 candidate metal-poor
stars selected from the RAVE database using a preliminary version of
the criteria described above.  The observations were carried out with
several telescope/spectrograph combinations,
including Magellan-Clay/MIKE, APO-3.5m/ARCES, AAT/UCLES, Max
Planck-2.2m/FEROS, CFHT/Espandons and VLT/UVES.  Full details of the
data aquisition and reduction are deferred to our paper on the derived
elemental abundances (Fulbright et al, in prep).  Briefly, all
spectrographs delivered a resolving power greater than 30,000 and,
with the exception of the UCLES set-up, the wavelength region covered
from below 4000 \AA{} to beyond 8000 \AA{}, albeit with some coverage
gaps. The effective wavelength range for the UCLES spectra is
4460--7260 \AA.  In each case the data were reduced using standard
methods for echelle data, utilizing pipeline 
programs when available.  The typical S/N level of the  
spectra is greater than 100/pixel and often exceeds 200/pixel.

The abundance analysis utilized Kurucz stellar atmospheres and the
MOOG program, with now the metallicity of the stellar atmosphere for
each star set equal to the value of [Fe~II/H] derived from the
analysis.  The value of the stellar effective temperature was derived
using the excitation temperature method based on \nion{Fe}{I} lines.
The mean difference of T$_{\rm ex}~-~$T$_{\rm phot}$ is $-66 \pm 138$~K.
The value of the stellar gravity was again taken to be that value
corresponding to the derived \teff{} on an old (12 Gyr) Yonsi-Yale
isochrone of the appropriate metallicity.  The microturbulent velocity
was set to the value that minimized the slope of the relationship
between the iron abundance derived from \nion{Fe}{I} lines and the
value of the reduced equivalent-width of each line.  The Fe~I lines
measured in the RAVE spectra are usually too weak in true VMP stars to
give reliable results.  The iron abundances derived from these lines
were useful, however, in rejecting from the analysis those (few) stars
for which the Ca Triplet lines appeared weak, but the Fe~I lines were
strong.  Further inspection of the spectra of these stars revealed
that the Ca lines are likely weakened by emission
cores.

The iron abundances from the echelle data provide an immediate check
on our selection criteria. The result is encouraging: of the 92 stars
with echelle data for which we predicted [Fe/H] $<-2$ from the RAVE
data, 87 (95\%) indeed have [Fe/H] values below $-2$.

\begin{figure}
\plotone{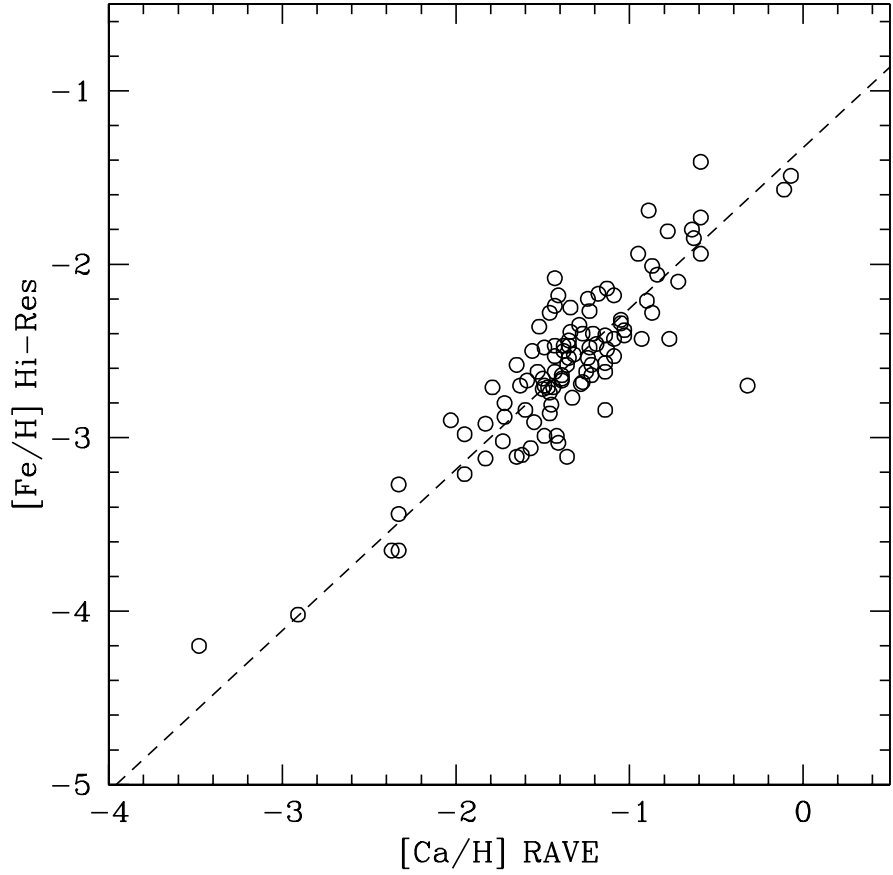}
\caption{Calibration of the calcium abundance derived from the RAVE
  spectra onto iron abundance, using [Fe/H] from high-resolution echelle spectra,
  for 112 stars.
   }
\label{calib}
\end{figure}

Calibration of the RAVE equivalent width abundances is then achieved
through least-squares fitting between the echelle-based iron
abundance, [Fe/H]$_{\rm{Hi-Res}}$, and the equivalent-width based
calcium abundance from the RAVE spectra, [Ca/H]$_{\rm{RAVE}}$.  The
correlation between these two quantities is shown in
Figure~\ref{calib} and the relationship is:
$$ {\rm [Fe/H]_{Hi-Res}} = 0.93 {\rm [Ca/H]_{RAVE}} -1.33,$$
with a correlation coefficient of 0.82, and standard deviation of the
residuals of 0.25~dex.

\section{The Stellar [Fe/H] Abundance Distribution Function at low metallicity}

\begin{figure}
\plotone{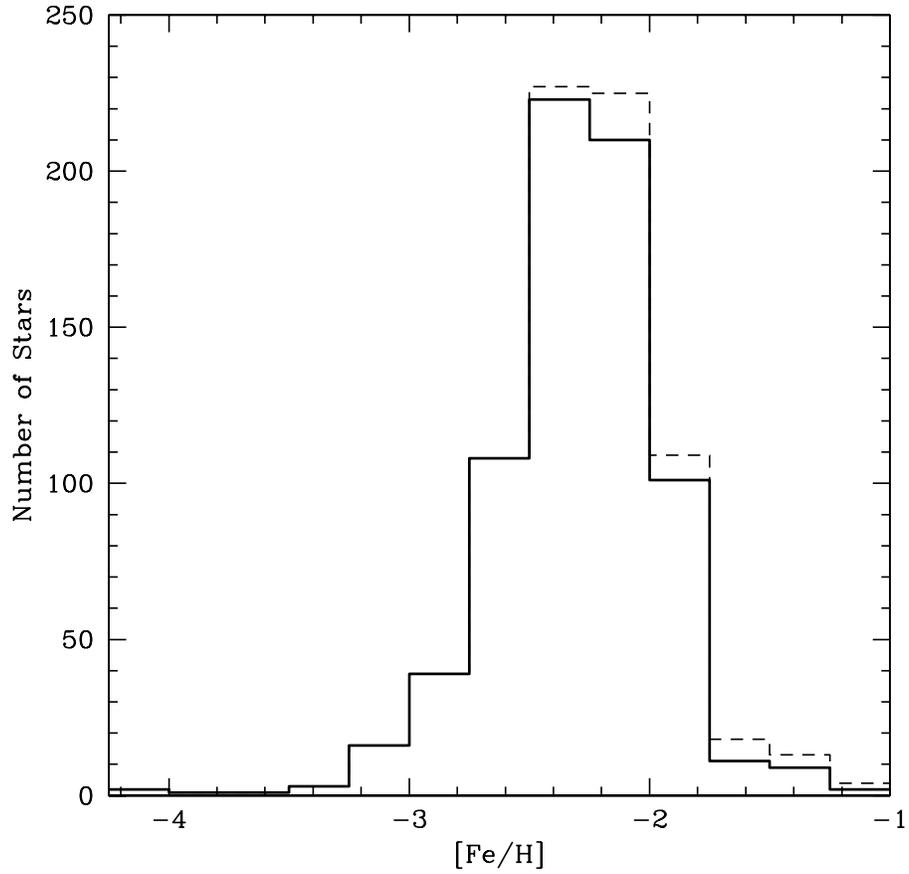}
\caption{Distribution of calibrated iron abundances [Fe/H], 
for the  sample of VMP stars selected from the RAVE database (solid line).
  The dashed histogram includes the VMP stars from the re-analysis of RAVE spectra for candidates from the HES survey and from 
  Ruchti et al.~(2010).}
\label{mdf}
\end{figure}

The calibration obtained from the echelle data was applied to the 771
VMP candidates from RAVE for which the MOOG analysis provided the
calcium abundance. The RAVE-based [Fe/H] abundance distribution
function that resulted is shown as the solid-line histogram in
Figure~\ref{mdf}; this contains a total of 612 stars with true iron
abundance [Fe/H] $ < -2$~dex.

\subsection{Distribution Function Completeness}

Given that our candidate sample had an upper metallicity cutoff determined by uncertain quantities, significant incompleteness at
the high-metallicity end is expected. We check this using published
catalogs of candidate VMP stars and a second RAVE study.

A SIMBAD-aided literature search showed that only 47 of our 612 very
metal-poor stars ([Fe/H] $< -2$~dex) have previously been proposed as
being metal-poor, from low-resolution spectroscopic
analyses, including 21 from \citet{frebel} and 11 from \citet{bon80}.
Ten of the 138 stars for which we derive $-2 <$ [Fe/H] $< -1$ also
have entries in SIMBAD identifying them as VMP
candidates.  

The most uniform comparison is with the Hamburg-ESO survey (HES) catalog of over 20,000 candidate VMP
stars  \citep{c08}. 
Cross-identification with  the RAVE database (at that time with 200,000 entries) yielded  473 matches, with HES metallicity
estimates for 296 stars, all with [Fe/H]$_{\rm HES} \leq -2$.  109 of
these stars are included in our RAVE VMP sample.  The RAVE pipeline
gave  \mhr $< -1$ for 76 of the remaining 187 stars.  We
reanalyzed these RAVE spectra as above, obtaining [Fe/H] $< -1$ for only 22
stars, with just 4 with [Fe/H] $< -2$, of which the most metal-poor
has [Fe/H] $= -2.26$.  These stars are therefore all very close to the
VMP threshold.

The high-resolution sample of \citet{greg} was selected from the same
RAVE catalog as this study and includes 20 stars for which those
authors derive [Fe/H] $< -2$ from their echelle spectra but which were
not selected by our VMP criteria. We re-analyzed the RAVE spectra of
these stars following the procedures developed above and found good agreement: [Fe/H] $<-2.0$ for 15 stars, and all 20 stars having [Fe/H]
$<-1.7$.  The lowest iron abundance of this group is
$-2.41$~dex, within 2$\sigma$ of our calibration.  Approximately half
of these stars have [m/H]$_{{\rm RAVE}}$ between $-1.5$ and $-1.8$,
the remainder having RAVE spectra of low quality.  As expected,
measuring errors in our parent RAVE sample lead to increasing
incompleteness in our final sample, at above  [Fe/H]$\sim
-2.5$.

\subsection{The low metallicity tail of the distribution function}

At very low metallicities, we test for true VMP stars observed by RAVE, 
but excluded from our sample, by cross-matching the RAVE database with
published catalogs of  confirmed VMP stars. There are 253
RAVE stars in the HERES project \citep{bar05}, which provides
high-resolution follow-up for HES catalog stars.  There
are only two candidate VMP matches: J112243.4-020936 (RAVE) is HE1120-0153
(HERES), and J132244.1-135531 (RAVE) is HE1320-1339 (HERES).  Our VMP
star criteria identified both, and our calibration gave
each an iron abundance of [Fe/H] $= -2.6$.  The HERES values are
[Fe/H] $ = -2.77 $ and $ -2.78$, respectively, in excellent agreement. No
confirmed very metal-poor star has been excluded from our RAVE sample.

There are three stars in our RAVE sample at or below
$-4$~dex.  These are C0022448-172429 for which we derive, from this
calibration, [Fe/H] $ = -4.0$~dex, CD~$-38^\circ245$ for which we derive
[Fe/H] $ = -4.2$~dex, and a third star, with a RAVE spectrum at our
minimum acceptable RAVE
S/N cutoff, which our calibration gives [Fe/H] $ = -4.0$~dex.   Our
echelle-based result for C0022448-172429 is [Fe/H]$_{\rm{Hi-Res}} =
-4.02$, and for CD~$-38^\circ245$ is [Fe/H]$_{\rm{Hi-Res}} = -4.20$. CD~$-38^\circ245$ was previously confirmed from echelle data to
have [Fe/H] $= -4.2$ by \citet{cay04}.   Our RAVE-based calibration is 
indeed providing true iron
abundances to $\sim 0.25$~dex accuracy, even at the lowest known
metallicities.

Our analysis is the first confirmation of the `ultra metal-poor' nature, 
[Fe/H] $< -4$~dex, of C0022448-172429. The low resolution,
objective prism HES spectrum for C0022448-172429 (Christlieb et
al.~2008, where this star is identified as HE0020-1741) provided
estimates of $-3.0$ and $-3.3$, depending on which method those
authors used. Christlieb et al.~further estimate that C0022448-172429
is carbon-rich, with [C/Fe] $= +1.0$.  Our echelle analysis confirms
the carbon-rich nature, with our value being [C/Fe] = $+0.9$, and we
find this star has an extremely high nitrogen abundance, [N/Fe] =
$+3.2$. Thus this star joins the select group of  carbon-enhanced metal poor (CEMP) stars \citep[see][for  a recent discussion of this class of object]{JEN10}. 

Although this makes our selection function not internally consistent,
adding the identified extra VMP stars  after re-analysis of their RAVE spectra (15 from Ruchti et al., plus 4
from the HES cross-check)   to our main
sample of 612 VMP stars, results in the dashed histogram in Fig.~\ref{mdf}. 
This  contains 631 stars with [Fe/H] $\le -2$~dex,
i.e~`Very Metal-Poor', and two stars with [Fe/H]$ < -4$, i.e. `Ultra
Metal-Poor' and one at $\sim -4$~dex. 567 of these VMP stars were not
known to be VMP previously, and one of the  two UMP stars is a new
confirmation.  The remaining star at $\sim -4$~dex has no previous published abundance determination and due to its marginal quality RAVE spectrum  remains as a candidate, awaiting 
scheduled high resolution
follow-up; if confirmed this would be a new
discovery.

\section{Conclusions}

The RAVE Survey allows identification of  large and complete
samples of apparently bright, very metal poor (VMP) stars, 
ideal for subsequent detailed analysis.  We have defined a sample of
612 stars with [Fe/H] $<-2$, based on some 200,000 spectra in the RAVE
database (which currently contains 350,000 spectra). This includes two
stars with [Fe/H] $ \le -4$~dex, one of which was previously known,
the other of which is a new ultra metal-poor star, with the previous
estimate of iron abundance being too high by $0.7$ to $1.0$~dex. This
star belongs to the interesting carbon-enhanced metal  poor 
class, having [C/Fe] = $+0.9$, and [N/Fe] = $+3.2$.  A third (new) 
metallicity $-4$ candidate remains to be confirmed. Only six
\citep{fre10} such ultra metal-poor stars were known prior to this
study. Comparison to other samples indicates that our completeness
limit at the metal-rich end is [Fe/H] $\approx -2.5$. It is very
likely our sample is complete for more metal-poor stars.

We note here, and will discuss elsewhere, that the shape of our
metallicity distribution function below [Fe/H]$\sim -2.8$, where we
are complete, is in excellent agreement with the independent
determinations from the  (low-resolution) HES survey  \citep{Schorck09,Li2010}.

Our technique uses a selection criterion based on the values of
[m/H]$_{{\rm RAVE}}$ and effective temperature from the RAVE pipeline
analysis, plus a photometric effective temperature. The original RAVE
spectrum of a star that passes the criterion is re-analysed to
determine the equivalent widths of the Ca T lines and Fe~I lines. We
calibrate the (re-measured) RAVE metallicity onto true iron abundance,
[Fe/H], through an echelle-based calibration derived in this
paper. The efficiency of our technique for finding VMP stars is very
high, demonstrated by the fact that $\sim 95$\% of the stars that we
determine by our calibration to have [Fe/H]$ <-2$ indeed have
echelle-based [Fe/H] in this range.  There is no need for follow-up,
higher spectral resolution data to confirm this derived iron
abundance.  Since the mean I magnitude of the RAVE sample of VMP stars
is $\sim 10.5$, follow-up detailed elemental abundance analyses are
both straightforward and efficient. The RAVE survey is on-going, and
the final RAVE database should at least triple the current sample size.

{\it Facilities:} \facility{UKST (6dF)}, \facility{ARC (ARCES)},
\facility{CFHT (Espandons)}, \facility{Magellan:Clay (MIKE)},
\facility{VLT:Kueyen (UVES)}, \facility{Max (FEROS)}

\acknowledgements We thank the staff of Siding Spring Observatory,
Apache Point Observatory, and Las Campanas Observatory for their
assistance.  JPF thanks Jennifer Johnson and Howard Bond for 
useful suggestions. JPF, RFGW and GRR acknowledge support through
grants from the W. M. Keck Foundation and the Gordon and Betty Moore
Foundation, to establish a program of data-intensive science at the
Johns Hopkins University.  This publication made use of 2MASS, a joint
project of the University of Massachusetts and IPAC/Caltech, funded by
NASA and the NSF, and of the VizieR databses, operated at CDS,
Strasbourg, France. RFGW and GRR acknowledge support from the NSF
through grant AST-0908326.  Support for RAVE (www.rave-survey.org) has
been provided by institutions of the RAVE participants and by their
national funding agencies.  RFGW thanks the Aspen Center for Physics
for providing a stimulating environment during the writing of this
paper.

\clearpage
\epsscale{0.75}
\clearpage
\epsscale{0.75}
\clearpage
\epsscale{0.75}
\clearpage


\begin{thebibliography}{}

\bibitem[Alonso et al.(1999)]{alo99} Alonso, A., Arribas, S., \& Martinez-Roger, C. 1999, A\&AS, 140, 261
\bibitem[Aoki et al.(2007)]{aok07} Aoki, W., Beers, T. C., Christlieb, N., Norris, J. E., Ryan, S. G. \& Tsangarides S. 2007, \apj, 655, 492
\bibitem[Barklem et al.(2005)]{bar05}Barklem, P.S. et al. 2005, \aap, 439, 129
\bibitem[Beers \& Christlieb(2005)]{bee05}Beers, T. C. \& Christlieb, N. 2005, \araa, 43, 531
\bibitem[Bromm \& Larson(2004)]{bro04} Bromm, V. \& Larson, R.B.~2004, \araa, 42, 79
\bibitem[Bond(1980)]{bon80} Bond, H. E.  1980, \apjs, 44, 517
\bibitem[Cayrel et al.(2004)]{cay04} Cayrel, R., Depagne, E., Spite, M., Hill, V., Spite, F., Fran\c{c}ois, P., Plez, B., Beers, T., Primas, F., Andersen, J., Barbuy, B., Bonifacio, P., Molaro, P. \& Nordstr\"{o}m, B. 2004, \aap, 416, 1117

\bibitem[Christlieb et al.(2004)]{chr04} Christlieb N. et al.~2004, \aap, 428, 1027

\bibitem[Christlieb et al.(2008)]{c08} Christlieb N., Scharck T., Frebel A., Beers T.C., Wisotzki L. \& Reimers D. 2008, \aap, 484, 721 (HES)
\bibitem[Cutri et al.(2003)]{cut03} Cutri, R. M., et al. 2003, 2MASS All-Sky Catalog of Point Sources, VizieR, II/246
\bibitem[Demarque et al.(2004)]{yy} Demarque, P., Woo, J.-H., Kim, Y.-C. \& Yi, S. K. 2004, ApJS, 155, 667
\bibitem[Frebel et al.(2006)]{frebel} Frebel A., Christlieb N., Norris J.E., Beers T.C., Bessell M.S., Rhee J., Fechner C., Marsteller B., Rossi S., Thom C., Wisotzki L., Reimers D. 2006, \apj, 652, 1585
\bibitem[Frebel(2010)]{fre10} Frebel A. 2010, Astron.~Nach., 331, 474

\bibitem[Hog et al.(2000)]{hog} H{\oe}g E., Fabricius C., Makarov V. V., Urban S., Corbin T., Wycoff G., Bastian U., Schwekendiek P. \& Wicenec A. 2000, \aap, 355, L27
\bibitem[Li et al.(2010)]{Li2010} Li, H., Christlieb, N., Schorck, T.,
  Norris, J.E.,  et al. 2010, arXiv:1006.3985
\bibitem[Norris, Gilmore, Wyse, et al.(2010)]{JEN10} Norris, J.E.,
  Gilmore, G., Wyse, R.F.G.,  Yong, D. \& Frebel, A. 2010, \apj, 722, L104
\bibitem[Prantzos(2009)]{pra09} Prantzos, N. 2009, in `The Galaxy Disk in Cosmological Context', proc.~IAU Symposium 254, eds.~J. Andersen, J. Bland-Hawthorn, and B.  Nordstr\"{o}m (Cambridge: CUP), p.~381
\bibitem[Ruchti et al.(2010)]{greg} Ruchti, G. et al., 2010, \apj, 721, L92
\bibitem[Schorck et al.(2009)]{Schorck09} Schork, T., Christleib, N.,
  Cohen, J.,  Beers, T., et al. 2009, \aap, 507, 817
\bibitem[Sneden(1973)]{sne73} Sneden, C. 1973, \apj, 184, 839
\bibitem[Suda et al.(2008)]{suda} Suda, T., Katsuta, Y., Yamada, S., Suwa, T., Ishizuka, C., Komiya, Y., Sorai, K., Aikawa, M. \& Fujimoto, M. Y. 2008, PASJ, 60, 1159 
\bibitem[Schlegel et al.(1998)]{sch98} Schlegel, D.J., Finkbeiner, D.P. \& Davis, M. 1998, \apj, 500, 525
\bibitem[Starkenburg et al.(2010)]{sta10} Starkenburg, E., Hill, V., Tolstoy, E., González Hernández, J. I., Irwin, M., Helmi, A., Battaglia, G., Jablonka, P., Tafelmeyer, M., Shetrone, M., Venn, K., de Boer, T. 2010, \aap, 513, 34
\bibitem[Steinmetz et al.(2006)]{ste06} Steinmetz, M.  et al. 2006, \aj, 132, 1645
\bibitem[Zwitter et al.(2008)]{zwi08} Zwitter, T. et al., 2008, \aj, 146, 421
\end{thebibliography}
\end{document}